\documentclass[twocolumn,showpacs,preprintnumbers]{revtex4} 

\input{epsf}
\def\to{\rightarrow}
\def\gev{\mbox{GeV}}

\def\tev{\mbox{TeV}}

\def\MPL{{\it Mod. Phys. Lett.} }

\def\NP{{\it Nucl. Phys.} }
\def\PL{{\it Phys. Lett.} }
\def\PR{{\it Phys. Rev.} }
\def\PRL{{\it Phys. Rev. Lett.} }

\def\frac#1#2{{\textstyle{{#1}\over {#2}}}}

\def\lsim{\mathrel{\rlap{\lower4pt\hbox{\hskip1pt$\sim$}}
    \raise1pt\hbox{$<$}}}
\def\gsim{\mathrel{\rlap{\lower4pt\hbox{\hskip1pt$\sim$}}
    \raise1pt\hbox{$>$}}}
\def\sqr#1#2{{\vcenter{\vbox{\hrule height.#2pt
         \hbox{\vrule width.#2pt height#1pt \kern#1pt
         \vrule width.#2pt}
         \hrule height.#2pt}}}}

\def\beq{\begin{equation}}
\def\eeq{\end{equation}}
\def\beqa{\begin{eqnarray}}
\def\eeqa{\end{eqnarray}}

\begin{document}

\title{Supergravity  Inflation on the  Brane}

\author{M. C. Bento$^{1,2}$, O. Bertolami$^1$ and A.A. Sen$^3$}

\vskip 0.2cm

\affiliation{$^1$  Departamento de F\'\i sica, Instituto Superior T\'ecnico \\
Av. Rovisco Pais 1, 1049-001 Lisboa, Portugal}

\vskip 0.2cm

\affiliation{$^2$  Centro de F\'{\i}sica das 
Interac\c c\~oes Fundamentais, Instituto Superior T\'ecnico}

\vskip 0.2cm

\affiliation{$^3$  Centro Multidisciplinar de Astrof\'{\i}sica, 
Instituto Superior T\'ecnico}

\vskip 0.2cm

\affiliation{E-mail addresses: bento@sirius.ist.utl.pt; 
orfeu@cosmos.ist.utl.pt; anjan@x9.ist.utl.pt}

\vskip 0.5cm

\date{\today}

\begin{abstract}
We study $N=1$ Supergravity inflation in the context of the 
braneworld scenario. Particular attention is paid to the problem of the 
onset of inflation at sub-Planckian field values and the ensued 
inflationary observables.
We find that the so-called $\eta$-problem encountered in supergravity 
inspired inflationary models can be solved in the context of the braneworld 
scenario, for some range of the parameters involved. Furthermore, we obtain 
an upper  bound on the scale of the fifth dimension, $M_5 \lsim 10^{-3} M_P$, 
in case the  inflationary potential is quadratic in the inflaton field, 
$\phi$. If the inflationary potential is  cubic in $\phi$,  
consistency with observational data requires that 
$M_5 \simeq 9.2 \times 10^{-4} M_P$.

\vskip 0.5cm
 
\end{abstract}

\pacs{98.80.Cq, 98.65.Es \hspace{2cm}Preprint DF/IST-4.2002, 
FISIST/09-2002/CFIF}

\maketitle
\section{ Introduction}

In supersymmetric theories with a single supersymmetry generator
($N=1$), complex scalar fields are the lowest components, $\phi^a$, of
chiral superfields $\Phi^a$.  Masses
for fields will be generated by spontaneous symmetry breaking so that
the only fundamental mass scale is the reduced Planck mass, 
$M = M_P/\sqrt{8 \pi}$.  The $N=1$
supergravity theory describing the interaction of the chiral superfields, 
the scalar potential is given by 

\begin{equation}
V = {\rm e}^{K/M^2} \left[F^{A\dagger} (K_A^B)^{-1} F_B -
3 {|W|^2 \over M^2}\right] + {\rm D-terms}~~,
\label{V}
\end{equation}
where

\begin{equation}
F_A \equiv {\partial W \over \partial\Phi^A}
+ \left({\partial K \over \partial\Phi^A}\right) {W \over M^2},~~
K_A^B\equiv
{\partial^2 K \over \partial\Phi^A \partial\Phi_B^\dagger}~.
\end{equation}

\noindent
The K\"ahler function, $K$, sets the form of the kinetic energy terms of the
theory
while the superpotential, $W$, determines the non-gauge interactions. For
 canonical kinetic energy terms,
$K = \sum_A\phi_A^\dagger\phi^A$, the potential takes the relatively
simple form

\begin{equation}
V = \exp \left(\sum_A {\phi_A^\dagger \phi^A\over M^2}\right)
\left[\sum_B \left|{\partial W \over \partial \phi_B}\right|^2 - 3
{\left|W\right|^2 \over M^2}\right]~.
\label{pot}
\end{equation}

Expanding the (slowly varying) inflaton potential about the value of the 
inflaton field at which the quantum fluctuations of interest are produced 
giving rise to structure formation, $\tilde \phi$, by writing 
$\varphi ={\tilde \phi}+\phi$, we obtain a polynomial potential in $\phi$

\beq
V =\Delta^{4} \left[1 + c_1 {\phi\over M}+ c_2  \left(\phi\over M\right)^2 
+ \cdots\right]~~,
\label{eq:tpot}
\eeq
where we have factored out the mass parameter $\Delta$, which sets the 
scale of the inflationary phase and can be directly 
related with the amplitude of energy density perturbations.
In this work, we shall consider a class of models in which the evolution 
of the inflaton dynamics is controlled by a single power at the point 
where the observed density fluctuations are produced; in this case,  
the potential assumes the following form 

\beq
V \simeq \Delta^{4} \left[1 + c_n \left(\phi\over M\right)^n\right]~~.
\label{eq:gpot}
\eeq

The case where the second term is dominant and $n=2$, a situation 
typical of chaotic inflationary models, has already 
been analysed in Refs. \cite{Maartens,Bento1}, where it is shown that
specific features of the braneworld
scenario allow for current observational constraints
to be successfully accounted for with a single scale at the superpotential 
level; hence,   difficulties with higher order non-renormalizable
terms can be quite naturally avoided since it is possible to achieve 
successful inflation with sub-Planckian field values.

In this work, we consider the case where the first term is dominant and
$n=2$ or  $n=3$; regarding the former case,
it is well known that a generic supergravity theory gives contributions
of order $\pm H^2$ \cite{Bertolami} to the inflaton mass squared 
whereas inflation requires $\vert m^2 \vert \ll H^2$, this is  the so-called 
$\eta$ problem.
We will show that, in the braneworld scenario, this problem can be avoided, 
provided the $D=5$ Planck mass satisfies the condition 
$M_5 \lsim 10^{16}~\gev$.
This  kind of analysis has recently been done   for a particular
supergravity F-term hybrid inflation model \cite{JohnM}, namely the one of 
Ref. \cite{Dvali}. It is relevant to point out that hybrid inflationary 
models \cite{Linde,Cope} allow, as a result of the dynamics of two or more 
scalar fields, for successful realizations of the old inflationary 
type models, but have, in some instances, difficulties
in what concerns initial conditions. However, these problems are shown to be 
naturally solved in braneworld scenarios in the case where the potential is 
dominated by the mass term \cite{Mendes}.

We shall also consider the case where the quadratic term gets cancelled and,
therefore, the inflationary potential is cubic in $\phi$, as is the case
of the supergravity natural inflation model of Ref.~\cite{Adams}. We 
show that this model works in the braneworld scenario 
provided $M_5 \simeq 1.1 \times 10^{16}~\gev$.

\section{Requirements on the inflationary potential}

We shall consider the  five-dimensional brane scenario, where  one
 assumes that Einstein equations with a negative cosmological constant hold
 (an anti-De-Sitter space is required) in $D$-dimensions and that matter
fields are confined to the $3$-brane; then, the $4$-dimensional Einstein
 equation is given by \cite{Shiromizu}:

\beq 
G_{\mu \nu} = - \Lambda  g_{\mu \nu} + {8 \pi \over M_P^{2}} T_{\mu \nu} 
+ \left({8 \pi \over M_{5}^3}\right)^2 S_{\mu \nu} - E_{\mu \nu}~,
\label{eq:gmunu}
\eeq
where $T_{\mu \nu}$ is the energy-momentum on the brane, $S_{\mu \nu}$ is a 
tensor that contains  contributions that are quadratic in
$T_{\mu \nu}$ and $E_{\mu \nu}$ corresponds to the projection of the 
5-dimensional Weyl tensor on the 3-brane  (physically, for a perfect fluid, 
it is associated to non-local contributions to the pressure and energy 
flux). 
In a cosmological framework, where the 3-brane resembles our universe and the
 metric projected onto the brane is an homogeneous and isotropic flat
 Friedmann-Robertson-Walker (FRW) metric, the  Friedmann
 equation becomes
\cite{Shiromizu,Binetruy}:

\beq
H^2 = {\Lambda \over 3} +  \left({8 \pi \over 3 M_P^2}\right) \rho
+ \left({4 \pi \over 3 M_5^3}\right)^2 \rho^2 + {\epsilon \over a^4},
\label{eq:H2}
\eeq
where $\epsilon$ is an integration constant. The four and five-dimensional
 cosmological constants are related by

\beq
\Lambda = {4 \pi \over M_5^3} \left(\Lambda_5 + {4 \pi \over 3 M_5^3}~
\lambda^2 \right)~~,
\label{eq:Lam}
\eeq
where $\lambda$ is the 3-brane tension, and
the four and five-dimensional Planck scales through

\beq
M_P = \sqrt{{3 \over 4 \pi}} {M_5^3 \over \sqrt{\lambda}}~~.
\label{eq:MP}
\eeq

Assuming that, as required by observations, the cosmological constant 
is negligible in the early universe and since the last term 
in Eq. (\ref{eq:H2}) rapidly becomes unimportant after inflation sets in, 
the Friedmann equation becomes

\beqa
H^2& =& {8 \pi \over 3 M_P^2} \rho \left[1 + {\rho \over 2 \lambda}\right]\\
&=&\left( 4\pi\over 3 M_5^3\right)^2 \rho^2\left(1+{\rho_c\over \rho}\right)~,
\label{eq:H22}
\eeqa
with

\beq
\rho_c={3\over 2\pi}{ M_5^6\over M_P^2}~~.
\eeq

Hence, the new term in $\rho^2$ is dominant at high energies, compared
to $\lambda^{1/4}$, i.e. $\rho>\rho_c$, but quickly decays at lower energies,
and  the usual four-dimensional FRW cosmology is recovered.

Consistently, we shall assume that the scalar field is confined to the brane, so
that its field equation has the standard form

\beq
\ddot \phi + 3 H \dot \phi = -{d V\over d \phi}~.
\eeq

Consistency between the  slow-roll approximation  and  the
full evolution equations requires that there are  constraints on the slope and
curvature of the potential. One can define two slow-roll parameters
\cite{Maartens}

\begin{eqnarray}
\label{eq:epsilon}
\epsilon &\equiv&{M_{P}^2 \over 16\pi} \left( {V' \over V}
\right)^2  {1+{V/ \lambda}\over(2+{V/\lambda})^2}~~,\\
\label{eq:eta}
\eta &\equiv& {M_{P}^2 \over 8\pi} \left(
{V'' \over V} \right)  {1 \over 1+{V/ 2 \lambda}}~~.
\end{eqnarray}

Notice that both parameters are
suppressed by an extra factor $\lambda/V$ at high energies and that, at low
energies, $V\ll\lambda$, they reduce to the standard
form. The value of $\phi$ at the end of inflation can be obtained 
from the condition

\beq
{\rm max}\{\epsilon(\phi_F),|\eta(\phi_F)|\}= 1~~.
\label{eq:phif}
\eeq

The number of e-folds during inflation is given by $N =
\int_{t_{\rm i}}^{t_{\rm f}} H dt$, which becomes \cite{Maartens}

\beq
\label{eq:N}
N \simeq - {8\pi  \over M_{P}^2}\int_{\phi_{\rm i}}^{\phi_{\rm
f}}{V\over V'} \left[ 1+{V \over 2\lambda}\right]  d\phi~~,
\eeq
in the slow-roll approximation. We see that, as a result of the modification
in the Friedmann equation,
the expansion rate is increased, at high energies,  by a factor $V/2\lambda$.

The amplitude of scalar perturbations is given by \cite{Maartens}

\beq
\label{eq:As}
A_{s}^2 \simeq \left . \left({512\pi\over 75 M_P^6}\right) {V^3
\over V^{\prime2}}\left[ 1 + {V \over 2\lambda} \right]^3
\right|_{k=aH}~~,
\eeq
where the right-hand side should be evaluated as the comoving scale
equals the Hubble radius during inflation, $k=a H$. Thus the amplitude
of scalar perturbations is increased
relative to the standard result at a fixed value of $\phi$ for a given
potential.

The scale-dependence of the perturbations is described by the
spectral tilt \cite{Maartens}

\beq
n_{s} - 1 \equiv {d\ln A_{s}^2 \over d\ln k} \simeq -6\epsilon + 2\eta~~,
\label{eq:ns}
\eeq
where the slow-roll parameters are given in Eqs.~(\ref{eq:epsilon})
and~(\ref{eq:eta}). Notice that, as
$V/\lambda\to\infty$, the spectral index of scalar perturbations 
is driven towards the Harrison-Zel'dovich spectrum, $n_{s}\to 1$.

Finally, the ratio between the amplitude of tensor and scalar
perturbations is given by \cite{Langlois}

\beq
\left({A_{t} \over A_{s}}\right)^2 \simeq {3 M_P^2\over 16 \pi}
\left({V^\prime\over V} \right)^2 {2 \lambda\over V}~~.
\label{eq:r}
\eeq

In what follows we shall use the inflationary observables defined  to 
study two generic supergravity inflationary models of the type described
by Eq. (\ref{eq:gpot}). We should stress that we shall be interested in 
achieving a sucessful inflationary scenario with sub-Planckian field 
values so as to avoid the abovementioned problems with higher order 
non-renormalizable terms.

\section{Quadratic potential}

We first consider the case where the potential is 
quadratic in $\phi$ and we rewrite it as

\beq
V= V_0  + {1\over 2} m^2 \phi^2~~,
\label{eq:V2}
\eeq
and assume that the first term is dominant.

In supergravity, effective mass squared contributions of fields are given by 
\cite{Cope} 

\beq
{1\over 2} m^2= 8 \pi {V_0\over M_P^2} \approx 3 H^2
\label{scond}
\eeq
since the horizon of the inflationary De Sitter phase has a Hawking 
temperature given by $T_H = H/2\pi$ \cite{Bertolami}. 

Contributions like the ones of Eq. (\ref{scond}) lead to 
$\eta \equiv M_P^2 V^{\prime\prime}/8 \pi V \simeq 2$ ; however, the onset of 
inflation requires $\eta \ll 1$. Within the braneworld scenario, however, 
$\eta$ is modified, at high energies, by a factor
$\lambda/V$ (see Eq.~(\ref{eq:eta})). Hence, if the quantity,  
$\alpha \equiv V_0/ \lambda$, is sufficiently large, this problem 
is automatically solved by the brane corrections, as in this case, 
$\eta \simeq 4/ \alpha$ for large $\alpha$.

We shall now see that a constraint on $M_5$ can be obtained 
from the requirement that the magnitude of the energy density perturbations
ensuing from our inflationary setup explains the
anisotropies in the CMB radiation observed by COBE.

\begin{figure}[t]
\centering
\leavevmode \epsfysize=7cm \epsfbox{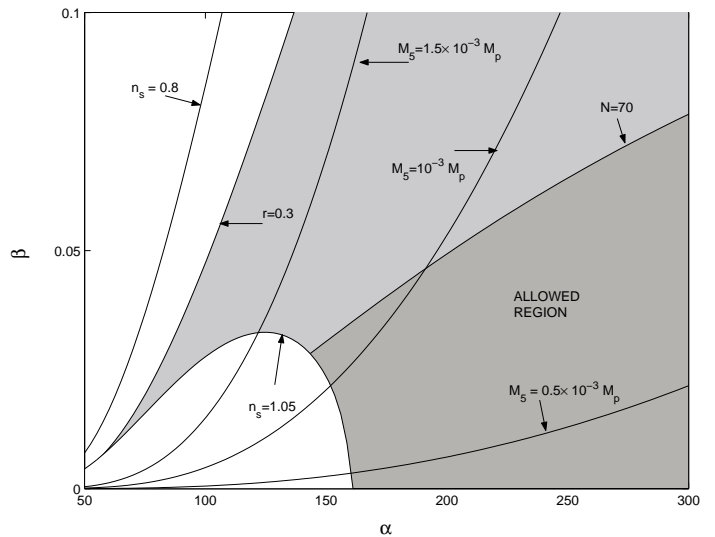}\\
\vskip 0.1cm
\caption{Contours of the inflationary observables  $n_s$ 
and $r$ in the $\alpha, \beta$ plane. The requirement that  the 
observational bounds  on these quantities (cf. Eq.(\ref{eq:observations})) 
be respected, leads to the shaded regions; after the condition that 
there is sufficient inflation is applied, $N\geq 70$, only the dark grey 
region remains allowed. We also show  contours corresponding 
to different values of $M_5$.}
\label{fig:figure1}
\end{figure}

The number of $e$-foldings, $N$, in terms of $\alpha$ is given by

\beq
N =  \alpha \left[{1\over 4} \log\left({\phi_I\over \phi_F}\right)
+ {2 \pi \over M_P^2}(\phi_I^2-\phi_F^2)
+ {16 \pi^2 \over M_P^4}(\phi_I^4-\phi_F^4)\right]~.
\label{eq:N2}
\eeq
Using the high energy approximation and $V \simeq V_0$ in 
Eq.~(\ref{eq:As}), we obtain for $A_s^2$

\beq
A_s^2 \simeq {64 \pi \over 75 M_P^6} {V_0^6 \over \lambda^3 m^4 \phi_\star}~~,
\label{eq:As2}
\eeq
where $\phi_\star$ is the value of $\phi$ when scales corresponding to
large-angle CMB anisotropies, as observed by COBE, left the
Hubble radius during inflation. For $N_\star\approx 55$,
$\phi_I=\phi_\star$ and $\phi_F= \beta M$ in 
Eq.~(\ref{eq:N2}), we get

\beq
\phi_\star \simeq 0.2 \beta \exp\left(220 \over \alpha\right) M_P~~,
\label{eq:phistar2}
\eeq
as, for sub-Planckian field values, 
the logarithmic term in Eq.~(\ref{eq:N2}) dominates. We prefer to leave 
$\phi_F$ as a free parameter since in hybrid models, which are of 
particular interest  because  they  can give rise to quadratic 
potentials of the type we are studying once some other field is held 
at the origin by its interaction with $\phi$,
inflation may end due to instabilities 
triggered by the dynamics of the other field. It then follows that the amount 
of inflation critically depends on the value of the inflaton field, 
$\phi_F$, and hence on $\beta$, 
at the time the instabilities arise. 
Notice that these 
instabilities are quite necessary in order to end inflation as 
$\epsilon = 64 \pi \phi^2/ \alpha M_P^2 << 1$ for $\alpha >> 1$  
and sub-Planckian field values. 
Moreover, it is clear that the condition $\epsilon \simeq 1$ cannot 
be met subsequently as the $\phi$ field is decreasing. 
   
We also mention that the problem of initial conditions for hybrid models, 
discussed in Ref. \cite{Mendes}, can  also be solved in this model 
due to the brane corrections.

Inserting Eq.~(\ref{eq:phistar2}) into 
Eq.~(\ref{eq:As2}) and using the fact that the  observed value
from COBE is $A_{s}=2\times 10^{-5}$, we obtain for 
$M_5$:

\beq
M_5^3 \simeq {8 \pi  \over \alpha^2}\times 10^{-5} \beta 
\exp(220/ \alpha)~M_P^3~~.
\label{eq:M5bound}
\eeq

Therefore, $M_5$ depends on the importance of the brane corrections via the 
$\alpha$ parameter and  the value of the field at the end of inflation 
through $\beta$.
In Figure \ref{fig:figure1}, we show contours
of the inflationary observables $n_s$ and $r \equiv 4 \pi 
\left({A_{t} \over A_{s}}\right)^2$, as given by Eqs. (\ref{eq:ns}) and 
(\ref{eq:r}), in terms of parameters 
$\alpha$ and $\beta$. The shaded regions in Figure 
\ref{fig:figure1} are obtained by requiring that  the bounds resulting from
latest CMB data from BOOMERANG \cite{Boom}, and MAXIMA \cite{Maxima} and 
DASI \cite{DASI}

\beq
0.8 < n_s < 1.05~~~, \qquad
r < 0.3~~,
\label{eq:observations}
\eeq
are satisfied. Also shown is the contour corresponding to $N=70$ (we have
chosen the initial value
of the inflaton field to be sub-Planckian, 
$\phi_I = 0.2~M_P$) and, after applying the condition that 
there is sufficient inflation,  $N \ge 70$,  only the darker 
region remains allowed. This analysis clearly leads to a constraint on 
parameter $\alpha$, $\alpha \gsim 140$.  
Finally, also in Figure \ref{fig:figure1}, we have  superposed 
contours of the scale $M_5$, from which we can derive an upper 
bound on this quantity

\beq
M_5 \lsim 10^{-3} M_P ~~.
\label{mass5}
\eeq

Finally, we mention that $\Delta$ (or instead $V_0$) 
in Eq. (\ref{eq:gpot}) can 
be estimated from the bound on the reheating temperature so as to avoid the 
gravitino problem (see Ref. \cite{Bento2} and references therein)

\beqa
T_{RH}& \lsim & 2 \times 10^{9}~\gev~~,~~ 6 \times 10^{9}~\gev~,\nonumber \\
\mbox{for}~~~~ m_{3/2} & = &1~\tev~~,~~10~\tev~~,
\label{eq:trhbound}
\eeqa 
leading to $\Delta \lsim 6.71 \times 10^{-4}~M_P$, 
as the reheating temperature is given 
by \cite{Ross}

\beq
 T_{RH} \simeq 5.5 \times 10^{-1} {\Delta^3\over M_P^2}~~.
\label{eq:trh}
\eeq

\section{Cubic potential}

We shall now consider the case where, due to some cancellation 
mechanism, the quadratic term is absent and the potential is cubic in 
$\phi$:

\beq
V=\Delta^4\left[1 + c_3 \left(\phi\over M\right)^3\right]~~.
\label{eq:V3}
\eeq
As mentioned before, we shall assume that the first term is dominant.
For instance, the model of Ref.~\cite{Adams} corresponds to precisely 
this case, with $c_3=-4$. We shall consider a generic $c_3$ and adapt our 
results for this particular example.

We start by computing the slow-roll parameters  $\epsilon$ and $\eta$:

\beq
\epsilon \simeq {18 c_3^2 \over \alpha} \left({\phi \over M}\right)^4~~, 
\label{eq:epsilon3}
\eeq

\beq
\eta \simeq - {12 c_3 \over \alpha} \left({\phi \over M}\right)~~.
\label{eq:eta3}
\eeq
where $\alpha\equiv {\Delta^4\over \lambda}$.

The value of $\phi$ at the end of inflation can be obtained from 
Eq.~(\ref{eq:phif}); we get, from $|\epsilon| \simeq 1$ 

\beq
\phi_F \simeq \left({\alpha \over 18 c_3^2}\right)^{1/4} M
\label{eq:phif3}
\eeq
while, from $|\eta| \simeq 1$, we obtain
\beq
\phi_F \simeq \left({\alpha \over 12 |c_3|}\right) M~~.
\label{eq:phif4}
\eeq
Hence, the prescription to be used depends on the value of $\alpha$. 
For $c_3 = - 4$, we see that the two prescriptions coincide 
for $\alpha \simeq 26$. 

The number of $e$-foldings, $N$, is given by:

\beq
N  =  {\alpha M \over 6 |c_3|} \left[{1 \over \phi_I} - 
{1 \over \phi_F}\right] = {\alpha M \over 6 |c_3|\phi_I} - 2~~,
\label{eq:N3}
\eeq 
using Eq. (\ref{eq:phif4}) (see below). 
Therefore, sufficient inflation
to solve the cosmological horizon/flatness problems, that is $N > 70$, 
is achieved for 

\beq
\phi_I < 2.3 \times 10^{-3} \left({\alpha \over |c_3|}\right) M~~,
\label{eq:phii5}
\eeq
yielding, for $\alpha = 26$ and $c_3 = - 4$:
\beq
\phi_I < 7.5 \times 10^{-2} M_P~~.
\label{eq:phii6}
\eeq

In order to match CMB anisotropies as observed by COBE, we compute 
the value of $\phi$, $\phi_\star$, that corresponds to $N = 55$:

\beq
\phi_\star = {\alpha M \over 342~|c_3|}~~.
\label{eq:phistar3}
\eeq

Inserting this result into
Eq.~(\ref{eq:As}), we obtain, in the high energy limit:

\begin{eqnarray}
A_s^2 & \simeq & {64 \pi \over 75 M_P^6} \alpha^3 \left({V \over V'}\right)^2 V
\nonumber \\
& = & {\alpha^4 \lambda \over 5400~\pi^2 c_3^2  \phi_\star^4}~~,
\label{eq:As3}
\end{eqnarray}
which determines $M_5$, using Eqs. (\ref{eq:MP}) and 
(\ref{eq:phistar3}):

\beq
M_5 = 7.3 \times 10^{-2} \left({A_s \over |c_3|}\right)^{1/3}M_P
= 9.2 \times 10^{-4} M_P~~,
\label{eq:valuem5}
\eeq
where we have set $|c_3| = 4$. Naturally, Eq. (\ref{eq:As3})
can also be used to extract the inflationary  scale, $\Delta$:

\beq
\Delta = 9.9 \times 10^{-4} \left({\alpha \over c_3^2}\right)^{1/4} M_P~~.
\label{eq:delta}
\eeq 

Requiring that the reheating temperature, computed using Eq.~(\ref{eq:trh}) 
and the above result, obeys the bounds necessary to avoid the gravitino 
problem, Eq.~(\ref{eq:trhbound}), leads, in turn, to the following 
bounds on $\alpha$ (for  $|c_3| = 4$) 

\beqa
 \alpha & \lsim & 3.4~~,~~14.7~,\nonumber \\
\mbox{for}~~ m_{3/2} &= &1~\tev~,~10~\tev~~.
\label{eq:alfboundd}
\eeqa 

\noindent
Notice that for such low values of $\alpha$, the relevant prescription for 
the end of inflation is the one given by Eq.~(\ref{eq:eta3}).
Moreover, we find that, within our approximations, 
$n_s$ does not depend on $\alpha$ and we get, for  $|c_3| = 4$,

\beq
n_s \simeq 0.93~~
\label{ns3}
\eeq
and
\beq
r = 2.5 \times 10^{-8} {\alpha^3 \over |c_3|^2} \simeq 5 \times 10^{-7}~~, 
\label{r3}
\eeq
where we have set $\alpha \simeq 14.7$; these values are clearly 
consistent with the observations. Notice that the observational limit on 
$r$ implies a relatively weak bound on $\alpha$, namely 
$\alpha < 577$.

\section{Conclusions}

In this work, we have analysed a class of $N=1$ supergravity inflationary 
models in which the evolution 
of the inflaton dynamics is controlled by a single power 
(quadratic or cubic in the inflaton field) at the point 
where the observed density fluctuations are produced, in the context 
of the braneworld scenario. 
We find that the so-called $\eta$-problem encountered in such 
models can be naturally solved in the context of braneworld 
scenario for some range of  $\alpha$, the ratio of the dominant term in the 
inflationary potential 
and the brane tension, and  the 
value of the inflaton field at the end of inflation, $\phi_F$.
For an inflationary potential with a quadratic term, we find that 
consistency with data requires $\alpha \gsim 140$. 
The implied mass scale of the fifth dimension is 
$M_5 \lsim 10^{-3} M_P$. For an inflationary potential that is cubic 
we find that, for consistency with observational data, 
$M_5 \simeq 9.2 \times 10^{-4} M_P$. Finally, we have checked that 
the gravitino problem  can be avoided  by a proper choice of the scale 
of the inflationary potential, $\Delta$, in both models.

\vfill
\begin{acknowledgments}

\noindent
M.C.B. and  O.B. acknowledge the partial support of Funda\c c\~ao para a 
Ci\^encia e a Tecnologia (Portugal)
under the grant POCTI/1999/FIS/36285. The work of A.A. Sen is fully 
financed by the same grant. 

\end{acknowledgments}

\end{document}